\documentclass[pra,preprint,showpacs,showkeys,preprintnumbers,amsmath,amssymb]{revtex4-1}

\usepackage{amsmath,amssymb,bm,epsfig}
\usepackage{dcolumn}
\usepackage{bm}
\usepackage{feynmf}

\begin{document}

\preprint{Phys. Rev. A {\bf 95}, 052110 (2017)}

\title{Infrared Problem in Quantum Acoustodynamics}

\author{Dennis P. Clougherty and Sanghita Sengupta}

\affiliation{
Department of Physics,
University of Vermont, 
Burlington, VT 05405-0125}

\date{\today}
\begin{abstract}
Quantum electrodynamics (QED) provides a highly accurate description of phenomena involving the interaction of atoms with light.  We argue that the quantum theory describing the interaction of cold atoms with a vibrating membrane--quantum acoustodynamics (QAD)--shares many issues and features  with QED.  Specifically, the adsorption of an atom on a vibrating membrane can be viewed as the counterpart to QED radiative electron capture.   A calculation of the adsorption rate to lowest-order in the atom-phonon coupling is finite; however, higher-order contributions suffer from an infrared problem mimicking the case of radiative capture in QED.  Terms in the perturbation series for the adsorption rate diverge as a result of massless particles in the model (flexural phonons of the membrane in QAD and photons in QED).  We treat this infrared problem in QAD explicitly to obtain finite results by regularizing with a low-frequency cutoff that corresponds to the inverse size of the membrane.  Using a coherent state basis for the soft phonon final state, we then sum the dominant contributions to derive a new formula for the multiphonon adsorption rate of atoms on the membrane that gives results that are finite, nonperturbative in the atom-phonon coupling, and consistent with the KLN theorem.  For micromembranes, we predict a reduction with increasing membrane size for the low-energy adsorption rate.  We discuss the relevance of this to the adsorption of a cold gas of atomic hydrogen on suspended graphene.

 \end{abstract}
\maketitle
\section{Introduction}

The adsorption process, where a free atom or molecule adheres to the surface of a material, is central to a variety of phenomena in surface science  \cite{cole-book}.  Experimental study of adsorption relies on having clean surfaces and gases at low temperature.  
 Recent experimental advances in the cooling and manipulation of ultracold atomic beams have opened up a new low-energy regime of adsorption where models of atom-surface interactions require a complete quantum mechanical treatment.  Furthermore, newly discovered materials such as graphene provide truly two-dimensional solids where strong low-frequency fluctuations of the surface might dramatically alter the adsorption dynamics \cite{dpc13}.

In addition to its centrality in surface science, it has been proposed that the dynamics of such a ``quantum hybrid'' system could be utilized for quantum information processing \cite{nanoEPR} or precision measurement \cite{optomech-review}.  It has been demonstrated that the interactions between a cold atom and a vibrating membrane can be engineered by placing the system in an optical cavity \cite{atom-membrane}.  Laser light can be bounced off the membrane to form an optical lattice \cite{atom-optomech} that can strongly couple over large distances the motion of the atoms to the vibrational modes of the membrane. 
The dynamics of cold atoms with nanotubes and cantilevered beams has also been studied both experimentally \cite{atom-nanotube, cobden-nanotube} and theoretically \cite{dpc03, cole-nanotube, nanotube-force}. 

Our focus here is the dynamics of a system consisting of a single atom interacting with an elastic membrane.  Such a system can be realized with a cold atom coupled via the van der Waals (vdW) interaction to a suspended two-dimensional material such as graphene.  While the vdW interaction between a cold neutral atom and graphene is weak, it is sufficiently strong for a hydrogen atom to bind to graphene at low temperatures.  We examine the transition rate of a cold atom to a bound state on the clamped membrane at zero temperature.  

It was previously recognized that there are similarities between the adsorption process and radiative capture \cite{dpc92,DPCPRB2016}.  Our work highlights the fact that quantum adsorption is a non-relativistic, condensed matter analogue of QED.  We essentially study in this work the phonon analogue to radiative corrections in scattering.  It is interesting to note that another well-known analogous effect, the phononic  Lamb shift,  has been detected in a recent experiment \cite{phonon-lamb} in a related system.  

A straightforward perturbative expansion of the adsorption rate in the atom-phonon interaction has terms that become infrared divergent for macroscopic membrane size.  In analogy to the well-known infrared problem in QED \cite{landaulifshitz4}, we show explicitly that the infrared divergences in the perturbation expansion of the adsorption rate can be remedied by using an appropriate set of soft phonon final states and summing over contributions corresponding to multiphonon emission.  We subsequently obtain a closed-form expression for the multiphonon adsorption rate.  The calculated rate is infrared finite, a result consistent with the KLN theorem \cite{kinoshita,LN}.


\section{Model}
\label{sec:model}
\begin{figure}
\includegraphics[width=12cm]{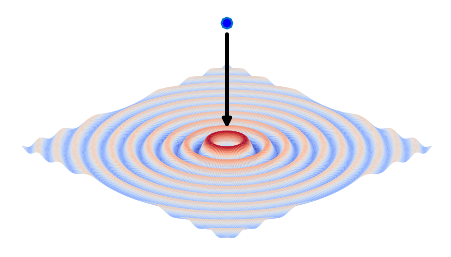}
\caption{\label{fig:model}  Atom impinging on an elastic membrane.  During adsorption, energy is transferred from the atom to the membrane and is radiated away by phonons.}
\end{figure}

We take for our model the following Hamiltonian $H=H_a+H_{ph}+H_{bi}+H_{ki}$ where
\begin{eqnarray}
H_a&=&E_k c_k^\dagger c_k -E_b b^\dagger b,\\
H_{ph}&=&\sum_q{\omega_q {a_q^\dagger} a_q},\\
H_{bi}&=&- g_{bb} b^\dagger b \sum_q  ({a_q+a_q^\dagger}), \\
H_{ki}&=&- g_{kb} (c_k^\dagger b+b^\dagger c_k) \sum_q  ({a_q+a_q^\dagger})
\label{ham}
\end{eqnarray}
The model was derived previously \cite{dpc13} by assuming that the atom moves slowly in the perpendicular direction toward an elastic, clamped membrane under tension and can transfer energy through force-coupling to the membrane by exciting its circularly symmetric flexural modes (Fig.~\ref{fig:model}).  The model has linear phonon dispersion with constant transverse speed of sound. 

Here, $c_{k}$ $(c_{k}^{\dagger})$ annihilates (creates) an atom in the continuum state with energy $E_{k}$;  $b$ $(b^{\dagger})$ annihilates (creates) an atom bound to the static membrane  with energy -$E_b$; $a_q$   $(a_q^{\dagger})$ annihilates (creates)  a circularly symmetric flexural phonon in the membrane with energy $\omega_q$;  $g_{kb}$ is the atom-phonon coupling for an atom in the continuum state; and $g_{bb}$ is the atom-phonon coupling for an atom bound to the membrane.  

If one regards the atom-phonon interactions as perturbations $H_i=H_{bi}+H_{ki}$,  the adsorption rate can be estimated using Fermi's golden rule.  The unperturbed Hamiltonian $H_0=H_a+H_{ph}$ has eigenstates $|n_k, n_b;\{n_q\}\rangle$ labeled by the number of atoms in the continuum $n_k$, the number of atoms bound $n_b$ and the number of phonons in each vibrational mode of the membrane $\{n_q\}$. The initial state has an atom in the continuum and the membrane in its ground state; the final state has a bound atom plus a phonon present.  The adsorption rate $\Gamma_0$ for a micromembrane is then 

\begin{eqnarray}
\Gamma_0&=& 2\pi\sum_q \delta(E_k+E_b-\omega_q) |\langle 0,1;1_q|H_i|1,0;0\rangle|^2\\
&=& 2\pi g_{kb}^2 \rho
\label{gr0}
\end{eqnarray}
where $\rho$ is a partial phonon density of states for the circularly symmetric vibrational modes of the membrane.  This result, finite in the limit of a large membrane, is of order $g_{kb}^2$ and is independent of $g_{bb}$.  Higher order contributions in $g_{bb}$ however are divergent in the large membrane limit where soft phonon emission becomes possible.  We use the inverse size of the membrane $\epsilon$ as an infrared regulator that provides a low frequency cutoff to the vibrational spectrum; $\omega_D$ is the membrane's high frequency limit.

The KLN theorem \cite{kinoshita,LN} informs us that infrared divergences are specious and are not contained in the true physical adsorption rate as $\epsilon\to 0$. Thus, approximations to the adsorption rate obtained by truncation of this (divergent) perturbation expansion must be carefully scrutinized for large membranes \cite{dpc12}.

If  one regroups the Hamiltonian so that $H_{bi}$ is included in the unperturbed Hamiltonian, a quite different result follows for the lowest order adsorption rate.  With $H_{ki}$ as the only perturbation, the remaining terms of $H$ form the unperturbed Hamiltonian which can be diagonalized with a canonical transformation (see Appendix).  We find that under the perturbation $H_{ki}$, the adsorption rate $\Gamma_1$  is given by
\begin{equation}
\label{1phrate}
\Gamma_1=2\pi g_{kb}^2 \rho e^{-2F} \bigg(1+{\Delta\over E_s}\bigg)^{-2}
\end{equation}
where $F= {g_{bb}^2\over 2}\sum_q {1\over\omega_q^2}$, $\Delta= {g_{bb}^2}\sum_q {1\over\omega_q}$, and $E_s=E_k+E_b$.   $\Gamma_1$ is of order $g_{kb}^2$ and contains effects of $g_{bb}$ to all orders.  Most importantly, in the large membrane limit ($\epsilon\to 0$), we see that $F\sim {\rho g_{bb}^2\over 2}\int_\epsilon {d\omega\over \omega^2}$, consequently growing as $1/\epsilon$.  Thus, the lowest order adsorption rate becomes exponentially small for large membranes, a result in dramatic contrast to Eq.~\ref{gr0}.  

The rate of adsorption producing a multiphonon final state is found in a similar fashion (see Appendix).  
In the large membrane regime, the rate of adsorption via emission of $n$ phonons  is 
\begin{equation}
\Gamma_n\approx 2\pi g_{kb}^2 \rho e^{-2F} \bigg(1+{\Delta\over E_s}\bigg)^{-2} \bigg({g_{bb}^2\rho\over\epsilon}\bigg)^{n-1}{1\over (n-1)!}
\label{Nphrate}
\end{equation}
We conclude that the adsorption rate for {\it any} finite number of phonons emitted is vanishingly small in the large membrane regime.  The situation here is reminiscent of bremsstrahlung emission by a charged particle in QED \cite{landaulifshitz4}.  We also note that for large membranes where $\epsilon\ll \rho g_{bb}^2$, the multiphonon rate exceeds the 1-phonon rate $\Gamma_1$, a familiar situation for soft photon emission from a scattered electron in QED.  

\begin{figure}
\includegraphics[width=12cm]{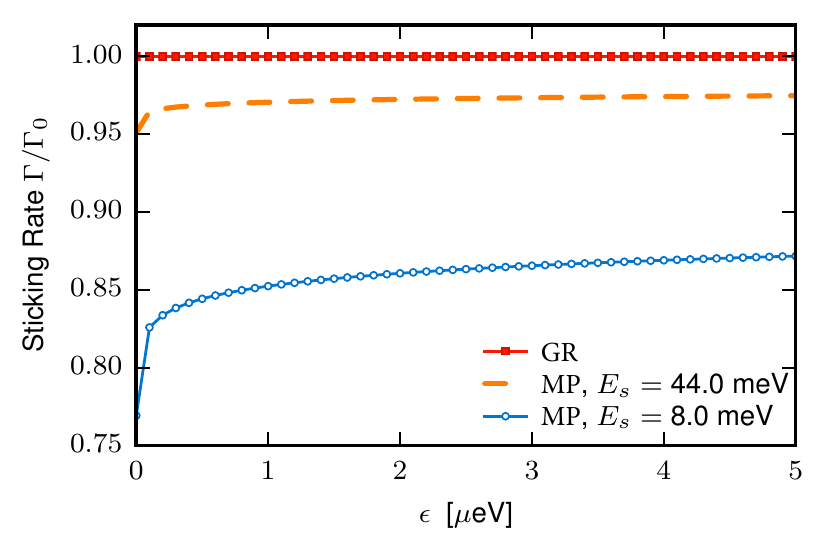}
\caption{\label{fig:eps}  Normalized sticking rate $\Gamma/\Gamma_0$ versus the IR cutoff $\epsilon$ for micromembranes with different binding energies.  Multiphonon processes (MP, Eq.~\ref{phrate}) suppress the simple golden rule (GR, Eq.~\ref{gr0}) adsorption rate $\Gamma_0$.  The suppression of the adsorption rate is stronger for shallow bound states.  GR can be a poor approximation for large membranes where soft phonon processes are important.}
\end{figure}

To obtain the total adsorption rate, we can sum over all $n$ and obtain a finite result; namely,
\begin{equation}
\Gamma=\sum_{n=1}^\infty \Gamma_n \approx 2\pi g_{kb}^2 \rho  \bigg(1+{\Delta\over E_s}\bigg)^{-2}
\label{phrate}
\end{equation}
We conclude that, as $\epsilon\to 0$, a non-vanishing adsorption rate results only with the emission of an infinite number of phonons.  We further note that the multiphonon rate differs from the simplest golden rule estimate $\Gamma_0$ by a fractional factor ${\cal R}=\big(1+{\Delta\over E_s}\big)^{-2}$ which depends logarithmically on the IR cutoff $\epsilon$ (see Fig.~\ref{fig:eps}).  Lastly, we observe that all adsorption rates $\Gamma_n$, as well as the total rate $\Gamma$, are finite (specifically, tending to zero) in the infrared limit $\epsilon\to 0$ in accord with the KLN theorem.

\section{Summary}
We have studied a quantum model for low-energy atomic adsorption on a 2D membrane.  The strength of the atom-phonon interaction at low frequencies gives rise to an infrared divergence problem in the adsorption rate for large membranes.  We use the inverse membrane size $\epsilon$ as a natural IR regulator.  With a canonical transformation, we obtain the adsorption rate in each sector of the phonon Fock space, and we show that each rate vanishes exponentially in the  limit of a vanishing cutoff $\epsilon\to 0$. 
We sum the dominant  terms in each sector of the phonon Fock space for small $\epsilon$ to obtain a new adsorption rate formula.  We show that the simplest golden rule approximation to the adsorption rate lacks a factor that reduces the adsorption rate, vanishing slowly as $\ln^{-2}(\omega_D/\epsilon)$ as $\epsilon\to 0$.  This prediction might be experimentally tested by comparing the cold atom adsorption rates by samples with membranes that range up to macroscopic size.

This result provides the answer to the question raised previously \cite{LJreply} concerning the effect of the IR cutoff on the low-energy sticking of atomic hydrogen on suspended graphene \cite{lepetit-jackson}. It was anticipated \cite{LJreply} that changes in the low-frequency phonon spectrum will have ``little impact'' on the sticking process.  This logic is used to justify an approximation of the phonon spectrum that ignores phonons with energies below $0.79$ meV.   Surprisingly, our result illustrates that the low-frequency phonon spectrum can have a substantial effect on the sticking, as the emission of an infinite number of soft phonons can dominate the sticking process for micromembranes. This differs from ``quantum sticking'' \cite{dpc92} where the emission of a finite number of quanta facilitates the adsorption.
Lastly, we note that this result is in  agreement with  previous results based on a variational mean-field method \cite{dpc10,dpc11,dpc13} that find a suppression of the adsorption rate with respect to the simple golden rule rate for sticking via emission of a finite number of phonons.


\section{Acknowledgments}
We gratefully acknowledge useful discussions with Professor V.N. Kotov on a variety of issues pertaining to this work.  

\appendix*
\section{}
We group the Hamiltonian in Eqs.~(1-4) as $H=H_0+H_1$ where
\begin{eqnarray}
H_0&=&E_k c_k^\dagger c_k -E_b b^\dagger b+\sum_q{\omega_q {a_q^\dagger} a_q}
- g_{bb} b^\dagger b \sum_q  ({a_q+a_q^\dagger}), \\
H_{1}&=&- g_{kb} (c_k^\dagger b+b^\dagger c_k) \sum_q  ({a_q+a_q^\dagger})
\end{eqnarray}
$H_0$ contains the independent boson model \cite{mahan} Hamiltonian and can be diagonalized with a canonical transformation
\begin{eqnarray}
{\tilde H_0}&=& e^S H_0 e^{-S}\\
&=& E_k c_k^\dagger c_k -(E_b+\Delta) b^\dagger b+\sum_q{\omega_q {a_q^\dagger} a_q}
\end{eqnarray}
where $S=-b^\dagger b\sum_q {g_{bb}\over\omega_q}(a_q^\dagger-a_q)$ and $\Delta= {g_{bb}^2}\sum_q {1\over\omega_q}$.  
The unperturbed ground state energy for one atom $E_g$ is thus $E_g=-(E_b+\Delta)$.

Since the 1-phonon eigenstate of ${\tilde H_0}$ is $|0,1;1_q\rangle$ with energy $E_q=-(E_b+\Delta)+\omega_q$, the corresponding unperturbed eigenstate of $H_0$ is $e^{-S} |0,1;1_q\rangle$, a product of phonon coherent states over the modes.  (Technically, one state in the product is a ``phonon-added'' coherent state.) Thus, from the golden rule, the adsorption rate for 1st order transitions under the perturbation $H_1$ is 
\begin{eqnarray}
\Gamma_1&=& 2\pi\sum_q \delta(E_k+E_b+\Delta-\omega_q) |\langle 0,1;1_q|e^S H_1|1,0;0\rangle|^2\\
&=&2\pi g_{kb}^2\sum_q \delta(E_k+E_b+\Delta-\omega_q) |\langle 1_q|X^\dagger\sum_{q'} a^\dagger_{q'}|0\rangle|^2
\end{eqnarray}
where $X=\exp(\sum_q {g_{bb}\over \omega_q}(a_q^\dagger-a_q))$.  We evaluate the phonon matrix element and find that
\begin{equation}
\sum_{q'} \langle 1_q|X^\dagger a^\dagger_{q'}|0\rangle=e^{-F} \bigg(1-{\Delta\over\omega_q}\bigg)
\end{equation}
where $F= {g_{bb}^2\over 2}\sum_q {1\over\omega_q^2}$.  Thus, 
\begin{eqnarray}
\Gamma_1&=& 2\pi g_{kb}^2 e^{-2F}\sum_q \bigg(1-{\Delta\over\omega_q}\bigg)^2 \delta(E_k+E_b+\Delta-\omega_q)\\
&\approx& 2\pi g_{kb}^2 e^{-2F}\int_\epsilon^{\omega_D} \rho d\omega \bigg(1-{\Delta\over\omega}\bigg)^2 \delta(E_k+E_b+\Delta-\omega)\\
&=& 2\pi g_{kb}^2 e^{-2F}\rho  \bigg(1-{\Delta\over E_s+\Delta}\bigg)^2 
\end{eqnarray}
where the quasi-continuum approximation is used to evaluate the sum.  As the IR cutoff $\epsilon$ approaches zero, $\Gamma_1$  exponentially vanishes.  

For final states obtained from 2-phonon eigenstates of ${\tilde H_0}$, the rate of adsorption is
 \begin{eqnarray}
\Gamma_2&=& 2\pi\sum_{\{n_{q}\} ,\  \sum_q n_{q}=2} \delta(E_k+E_b+\Delta-{\textstyle \sum_p n_{p}\omega_{p}}) |\langle 0,1;\{n_{q}\}| e^S H_1|1,0;0\rangle|^2\\
&\approx& 2\pi g_{kb}^2 e^{-2F}\rho^2 {g_{bb}^2\over\epsilon}  \bigg(1-{\Delta\over E_s+\Delta}\bigg)^2, \ \epsilon\to 0
\end{eqnarray}
Other contributions to $\Gamma_2$ are subdominant as $\epsilon\to 0$.  The use of a coherent state phonon basis for the final state removes the IR divergence.  This gives insight into the analogous method to remedy IR divergences in QED with coherent states \cite{IR-coherent}.  In our case, the coherent states are a natural phonon basis, given our choice of $H_0$.

The dominant contribution to the adsorption rate from $n$-phonon eigenstates has $(n-1)$ soft phonons ($\omega\sim \epsilon$) and a hard phonon ($\omega\sim E_s$) to satisfy energy conservation. 
 \begin{eqnarray}
\Gamma_n&=& 2\pi\sum_{\{n_{q}\} ,\  \sum_q n_{q}=n}  \delta(E_k+E_b+\Delta-{\textstyle \sum_p n_{p}\omega_{p}}) |\langle 0,1;\{n_{q}\}| e^S H_1|1,0;0\rangle|^2\\
&\approx& 2\pi g_{kb}^2 e^{-2F}{\rho^n\over (n-1)!} \bigg({g_{bb}^2\over\epsilon}\bigg)^{n-1}  \bigg(1-{\Delta\over E_s+\Delta}\bigg)^2, \ \epsilon\to 0
\end{eqnarray}
Thus, the IR divergences have been successfully removed to all orders in $g_{bb}$.

The total adsorption rate for a micromembrane is obtained by summing over the $n$-phonon contributions.
\begin{eqnarray}
\Gamma&=&\sum_{n=1}^\infty \Gamma_n \approx 2\pi g_{kb}^2 \rho e^{-2F} \bigg(1+{\Delta\over E_s}\bigg)^{-2}
 \sum_{n=0}^\infty \bigg({\rho g_{bb}^2\over\epsilon}\bigg)^n {1\over n!}\\
 &\approx& 2\pi g_{kb}^2 \rho  \bigg(1+{\Delta\over E_s}\bigg)^{-2}
\end{eqnarray}
since $2F=({\rho g_{bb}^2/\epsilon})$ in the quasi-continuum approximation for a dense phonon spectrum.  Remarkably, the exponentially decaying factor $e^{-2F}$ is cancelled with the infinite summation, and the resulting rate is the product of the simple GR rate $\Gamma_0$ with a cutoff-dependent reduction factor ${\cal R}=(1+{\Delta/ E_s})^{-2}$.  

\bibliographystyle{apsrev4-1}
\bibliography{qs,hybrid}

\end{document}